\documentclass[conference]{IEEEtran}
\IEEEoverridecommandlockouts
\usepackage{cite}
\usepackage{amsmath,amssymb,amsfonts}
\usepackage{algorithmic}
\usepackage{graphicx}
\usepackage{textcomp}
\usepackage{xcolor}
\usepackage{booktabs}
\usepackage{makecell}
\usepackage{mathtools}
\usepackage{url}
\def\BibTeX{{\rm B\kern-.05em{\sc i\kern-.025em b}\kern-.08em
    T\kern-.1667em\lower.7ex\hbox{E}\kern-.125emX}}

\newcommand{\metric}[1]{\textsf{\footnotesize #1}}

\begin{document}

\title{Decomposing Predictive Kubernetes Autoscaling for\\ Large Language Model Serving Under Long Startup Delays}

% \author{\IEEEauthorblockN{Anonymous Author(s)}
% \IEEEauthorblockA{\textit{Affiliation} \\
% \textit{Institution}\\
% City, Country \\
% email@example.com}
% }

\author{
\IEEEauthorblockN{Tianrui Liu}
\IEEEauthorblockA{
\textit{University of California, San Diego}\\
La Jolla, CA, USA\\
til028@ucsd.edu
}
\and
\IEEEauthorblockN{Xiaohai Hu}
\IEEEauthorblockA{
\textit{University of Washington}\\
Seattle, WA, USA\\
huxh@uw.edu
}
}

\maketitle

\begin{abstract}
Large language model (LLM) inference deployed on Kubernetes faces an autoscaling challenge that conventional web services do not: new serving replicas take two to ten minutes to start because multi-gigabyte model weights must be loaded, which makes purely reactive scaling structurally late. We ask a sharp question: among the components of a predictive autoscaler, which ones actually matter under such long actuation delays? We answer it by decomposing predictive autoscaling into four factors---token-aware demand tracking, startup-delay lookahead, a bounded uncertainty margin, and plant-state observation---and measuring each factor's contribution in isolation on production-derived heavy-tailed workloads generated by ServeGen. Our main finding is that a simple exponentially weighted moving average (EWMA) predictor with delay-aware lookahead and an upper confidence bound (UCB) margin captures most of the benefit, reducing time-to-first-token (TTFT) service-level-objective (SLO) violations from 53\% (reactive, queries-per-second based) to 0.5\% across five random seeds; lookahead alone is the single largest factor, a 14$\times$ reduction. Kalman filter variants do not consistently improve the cost--SLO tradeoff. Controlled experiments isolate the reason token granularity is necessary: context length, through key--value cache pressure, degrades TTFT far more than request rate at matched throughput. Finally, a validation on a real Kubernetes cluster (Qwen2.5-7B, A100, vLLM) confirms the central mechanism: a delay-aware lookahead controller cuts TTFT violations from 63.5\% to 3.7\% relative to reactive KEDA scaling. We distinguish throughout which findings are specific to Kubernetes actuation and which are general to LLM serving.
\end{abstract}


%% ============================================================
\section{Introduction}
%% ============================================================

Large language models (LLMs) such as GPT, Llama, and Qwen are increasingly deployed as online inference services on Kubernetes clusters. Autoscaling these services---adjusting the number of serving replicas to track offered load---is harder than autoscaling traditional web services for two distinct reasons. The first is \emph{general to LLM serving}, independent of the orchestrator; the second is \emph{specific to the Kubernetes actuation path}. We are careful to separate them throughout this paper, because conflating them has led prior work to attack the wrong part of the problem.

\textbf{General to LLM serving: load is not requests.} Each request processes a variable number of input tokens (the prompt) and generates a variable number of output tokens (the completion). Inference proceeds in two phases with different resource profiles. The \emph{prefill} phase processes all input tokens in one parallel forward pass; it is compute-bound and sets the time-to-first-token (TTFT). The \emph{decode} phase emits output tokens one at a time auto-regressively; it is memory-bandwidth-bound and sets the time-per-output-token (TPOT). During decode, the model holds the attention key and value tensors for every active sequence in GPU memory---the ``KV cache.'' A request with 4096 input tokens consumes roughly $16\times$ the prefill compute and KV-cache memory of one with 256 tokens, yet a queries-per-second (QPS) signal treats the two identically. The KV cache, not GPU compute, is frequently the binding capacity constraint: compute can sit idle while a full cache blocks new admissions.

\textbf{Specific to Kubernetes: actuation is slow.} Launching a new LLM serving pod requires pulling and loading model weights (14--140~GB for 7B--70B parameter models), initializing CUDA contexts, and capturing execution graphs. This takes two to ten minutes---orders of magnitude longer than the seconds it takes to start a stateless web pod. By the time a reactive controller observes overload and provisions capacity, the transient that triggered scaling has already caused sustained SLO violations. This is a \emph{delayed-actuation control problem}: any reactive policy is structurally late by the startup delay $\Delta$.

Together these properties mean the autoscaler must \emph{predict} future demand and act $\Delta$ ahead of time. But a predictive autoscaler has many moving parts---demand estimators, lookahead horizons, uncertainty margins, plant-state observers, and increasingly elaborate state estimators such as Kalman filters. \textbf{Which of these actually matter?} Should practitioners invest in sophisticated predictors, or does a simple recipe capture the benefit? This is the central question of the paper, and to our knowledge it has not been answered with a controlled factor study.

\textbf{Approach.} We implement a single predictive-autoscaling framework and instantiate it as a family of variants that differ in exactly one design factor at a time, so that each factor's contribution can be read off directly. We evaluate on production-derived heavy-tailed workloads in a calibrated event-driven simulator, complemented by a real-cluster validation of the central mechanism. We decompose the design into four factors:
\begin{enumerate}
\item \textbf{Token-aware demand tracking}: monitoring prefill/decode token rates rather than QPS. Reduces violations from 53\% to 21\%.
\item \textbf{Startup-delay lookahead}: predicting demand at $t+\Delta$ instead of reacting to the present. Reduces violations from 21\% to 1.4\%---the largest single-factor improvement.
\item \textbf{Bounded uncertainty margin}: scaling for the upper confidence bound of predicted demand rather than the point estimate. Reduces violations from 1.4\% to 0.5\%.
\item \textbf{Plant-state observation}: tracking queue depth and KV-cache pressure in addition to demand. Marginal SLO improvement at higher replica cost.
\end{enumerate}

\textbf{Contributions.}
(1) A controlled factor decomposition showing that \emph{delay-aware lookahead is the dominant factor} in predictive LLM autoscaling, with the UCB margin adding incremental robustness and more complex predictors (Kalman filters) yielding diminishing or negative returns (Section~\ref{sec:eval}).
(2) System identification and controlled experiments establishing that \emph{token granularity is necessary because context length, via KV-cache pressure, dominates TTFT degradation}, while QPS alone is insufficient (Section~\ref{sec:sysid}).
(3) A real-Kubernetes validation of the central timing mechanism: predictive lookahead reduces TTFT violations from 63.5\% to 3.7\% versus reactive KEDA (Section~\ref{sec:hw}).
(4) An explicit accounting of which conclusions are Kubernetes-specific, which are general to LLM serving, and which depend on relative (not absolute) simulator fidelity.

%% ============================================================
\section{Background and System Model}
%% ============================================================

\subsection{LLM Serving Systems}

State-of-the-art serving systems such as vLLM~\cite{kwon2023vllm} and SGLang~\cite{zheng2024sglang} employ \emph{continuous batching}~\cite{yu2022orca}, dynamically admitting and retiring requests at each decode iteration rather than waiting for a batch to finish. The KV cache is managed via \emph{PagedAttention}~\cite{kwon2023vllm} in fixed-size blocks. Crucially, vLLM gates new request admission on KV-cache \emph{block availability} ($\texttt{free\_blocks}\ge\texttt{watermark}$), not on compute utilization: when free blocks fall below the watermark, arriving requests wait regardless of how idle the GPU is. This cache-gated admission rule is the architectural fact behind our system-identification result (Section~\ref{sec:sysid}) and is general to LLM serving, not specific to Kubernetes.

\subsection{Kubernetes Autoscaling}

The Kubernetes Horizontal Pod Autoscaler (HPA)~\cite{k8shpa} is a proportional controller, $\text{desired}=\lceil \text{ready}\times \text{metric}/\text{target}\rceil$, with a tolerance band and configurable stabilization windows. KEDA~\cite{keda2024} extends HPA with external metric sources such as a Prometheus query on queue depth. Both are reactive and orchestrator-generic: neither distinguishes token-heavy from token-light requests at equal QPS (an LLM-serving gap), and neither compensates for the multi-minute pod startup delay (the Kubernetes-specific gap). Our predictive framework addresses both, but our factor study shows the second gap is the one that dominates outcomes.

\subsection{System Model}
\label{sec:model}

We make the serving topology and its assumptions explicit, since the autoscaler's capacity arithmetic depends on them.

\textbf{Replicas.} The deployment consists of $R$ \emph{homogeneous} replicas, each a full single-GPU copy of the model ($\texttt{tensor-parallel-size}{=}1$). Prefill and decode are \emph{colocated} within a replica (continuous batching over a shared KV cache), not disaggregated across separate pools as in DistServe~\cite{zhong2024distserve} or Splitwise~\cite{patel2024splitwise}; the disaggregated case is out of scope and noted as future work.

\textbf{Load balancing.} A request is dispatched to the ready replica with the shortest queue (join-shortest-queue). This keeps per-replica load approximately balanced, which is what licenses the aggregate-demand-divided-by-per-replica-capacity arithmetic below.

\textbf{Capacity constants.} Each replica sustains a prefill throughput $C_p$ and a decode throughput $C_d$ (tokens/s). We estimate $C_p{=}8000$ and $C_d{=}500$ tokens/s per replica by profiling the simulator at its saturation point and matching it to Vidur's~\cite{agrawal2024vidur} hardware-calibrated capacity wall ($\sim$20~QPS per Llama-2-7B replica on an A100). These constants are deployment-specific inputs, not tuned per experiment; Section~\ref{sec:eval} reports sensitivity to the resulting operating point via a parameter sweep.

\textbf{Control variables.} The autoscaler observes, at a polling interval $T$, the aggregate prefill token rate $p_t$, decode token rate $d_t$, queue depth $q_t$, and KV-cache utilization $k_t$, all exported by vLLM as Prometheus metrics. It actuates a single integer, the replica count $R$, subject to a startup delay $\Delta$ on scale-up.

%% ============================================================
\section{Predictive Autoscaling Framework}
%% ============================================================
\label{sec:method}

All variants share a four-step structure; they differ only in how demand is estimated and what state is observed, which is what lets us attribute outcomes to individual factors.

\textbf{Step 1: Demand estimation.} At each interval ($T{=}15$\,s) estimate $p_t$ and $d_t$ from vLLM's \metric{prompt\_tokens\_total} and \metric{generation\_tokens\_total} counters.

\textbf{Step 2: Delay-aware lookahead.} Predict demand at $t+\Delta$ (the startup delay, typically 180\,s for a 7B model on an A100). Using a backward-difference slope,
\begin{equation}
\hat{\dot{p}}_t=\frac{\hat{p}_t-\hat{p}_{t-T}}{T},\qquad
\hat{p}_{t+\Delta}=\hat{p}_t+\hat{\dot{p}}_t\,(\Delta-\tfrac{T}{2}).
\end{equation}
The $\Delta-T/2$ correction accounts for the slope being centered at $t-T/2$. For the first two intervals ($t<2T$) the slope is undefined; we set $\hat{\dot{p}}_t{=}0$ and rely on the reactive floor $\hat{p}_t$ plus the margin below.

\textbf{Step 3: Bounded uncertainty margin.} Scale for an upper confidence bound rather than the point estimate. For each demand dimension $i\in\{p,d\}$,
\begin{equation}
d_i^{\text{ucb}}=\hat{x}_{t+\Delta,i}+\beta\,\sigma_{\text{eff},i},\quad
\sigma_{\text{eff},i}=\min(\sigma_i,\gamma\,|\hat{x}_{t+\Delta,i}|),
\end{equation}
with $\beta{=}1.5$ and $\gamma{=}0.8$ unless noted. The cap is essential: without it, propagated uncertainty grows unboundedly with the horizon, causing pathological over-provisioning at long $\Delta$. EWMA-UCB derives $\sigma_i$ from the EWMA of squared prediction errors; Kalman variants derive it from the covariance propagated over the horizon.

\textbf{Step 4: Replica computation.} Convert demand to replicas via the capacity constants of Section~\ref{sec:model}:
\begin{equation}
R^\star=\max\!\left(\left\lceil\frac{d_p^{\text{ucb}}}{C_p}\right\rceil,\ \left\lceil\frac{d_d^{\text{ucb}}}{C_d}\right\rceil,\ R_{\min}\right).
\end{equation}
Scale-up is applied immediately; scale-down uses a 300\,s stabilization window following HPA convention.

\subsection{Variants}

The variants below each add exactly one factor relative to the previous, isolating its effect.

\emph{Token-Reactive}: tracks $p_t,d_t$ via EWMA but scales on \emph{current} smoothed demand, with no lookahead and no margin. Isolates token awareness from prediction.

\emph{EWMA}: adds delay-aware lookahead (predicts at $t+\Delta$); demand only, no margin. The simplest token-aware predictive autoscaler.

\emph{EWMA-UCB}: adds the bounded margin. Maintains $\hat{p}_t=\alpha p_t+(1{-}\alpha)\hat{p}_{t-1}$ ($\alpha{=}0.3$), slope by consecutive differences, lookahead via Step~2, and $\sigma$ from the EWMA of squared errors. Despite its simplicity it captures most of the benefit.

\emph{KF-4state}: replaces the EWMA estimator with a Kalman filter~\cite{bar2004estimation} over $\mathbf{x}=[p_t,\dot{p}_t,d_t,\dot{d}_t]^\top$ using a constant-velocity model (per stream $\mathbf{F}_{\text{s}}=[\begin{smallmatrix}1&T\\0&1\end{smallmatrix}]$, $\mathbf{H}_{\text{s}}=[\begin{smallmatrix}1&0\end{smallmatrix}]$, block-diagonal over prefill/decode). It adapts trust between observation and trend via the Kalman gain and yields a calibrated $\sigma$ for the margin.

\emph{KF-6state}: extends the state with queue $q_t$ and KV pressure $k_t$, with transition rows coupling queue growth to prefill demand (gain $a_q$, drain $b_q$) and KV growth to decode demand ($a_k$, $b_k$), so demand uncertainty propagates into plant-state predictions.

\emph{Reactive-Plant}: observes $q_t,k_t$ but does not predict ahead; scales when thresholds are crossed. A Chiron-style~\cite{patke2025chiron} backpressure approach without delay compensation.

%% ============================================================
\section{Evaluation}
%% ============================================================
\label{sec:eval}

\subsection{Experimental Setup}

\textbf{Simulator.} We build an event-driven simulator (SimPy) modeling continuous batching, chunked prefill, and KV-cache block tracking, with sub-linear batch scaling $t_{\text{iter}}(b)=10+1.89\sqrt{b-1}$\,ms calibrated against Vidur~\cite{agrawal2024vidur} (validated to within 9\% of real A100 systems). Our simulator matches Vidur's capacity wall ($\sim$20~QPS per Llama-2-7B replica) and TPOT scaling (within $1.3\times$) but over-estimates \emph{absolute} TTFT by roughly $2\times$ due to simplified scheduling. We therefore restrict all simulator-based claims to \emph{relative} comparisons between policies; absolute latency values are not used as evidence. The real-cluster validation (Section~\ref{sec:hw}) covers the absolute regime.

\textbf{Workloads.} We generate production-representative workloads with ServeGen~\cite{xiang2025servegen}, fit to Alibaba's Bailian platform (3.5B requests over four months). Input lengths follow Pareto/log-normal distributions (mean 622, p95 1649, capped at 4096); outputs follow an exponential (mean 244, p50 66)---substantially heavier-tailed than Gaussian synthetic loads. Arrivals follow a burst pattern (5$\to$15~req/s) with Gamma inter-arrivals. We also use a synthetic burst trace (Gaussian 256/128-token lengths) for the cold-start sweep.

\textbf{Baselines and their tuning.} HPA-QPS (8~req/s target per replica), KEDA-Queue (threshold 5 waiting requests per replica), and Static($N$) (a fixed allocation chosen \emph{after} seeing the workload---a hindsight lower bound, not an online policy). All dynamic methods start at 2 replicas; default $\Delta{=}180$\,s. We do not claim the reactive thresholds are optimal. We swept neighboring thresholds for HPA and KEDA and the qualitative conclusion was unchanged: no reactive configuration reached the low-violation region without approaching static over-provisioning, because the limitation is structural (actuation latency), not threshold choice. The Pareto analysis below sweeps the predictive parameters so that no single configuration carries the argument, and Static($N$) provides the matched-cost over-provisioning reference the reactive baselines cannot reach online.

\textbf{Metrics.} TTFT SLO violation rate (TTFT${>}2000$\,ms among completed requests), average replica count (proportional to GPU cost), and a conservative failure rate (violations + dropped + unfinished, over total). We report mean$\pm$std over five seeds.

\subsection{Factor Decomposition}

\begin{table}[t]
\caption{Factor decomposition on the ServeGen workload (5 seeds, mean$\pm$std). Each row adds one factor relative to the row above; Kalman variants swap in a more complex estimator.}
\label{tab:decompose}
\centering
\small
\begin{tabular}{@{}lrrl@{}}
\toprule
Policy & Viol\% & Repl & Design factor \\
\midrule
HPA-QPS & 52.5{\scriptsize$\pm$3.1} & 2.1 & (reactive, QPS) \\
Token-Reactive & 20.9{\scriptsize$\pm$1.2} & 4.8 & + Token demand \\
EWMA & 1.44{\scriptsize$\pm$.68} & 9.9 & + Lookahead \\
EWMA-UCB & 0.54{\scriptsize$\pm$.32} & 9.5 & + UCB margin \\
KF-4state & 5.91{\scriptsize$\pm$6.9} & 9.2 & + KF estimation \\
KF-6state & 0.47{\scriptsize$\pm$.30} & 11.1 & + Plant state \\
Reactive-Plant & 19.0{\scriptsize$\pm$1.2} & 6.0 & (react to $q$/KV) \\
\midrule
Static(8) & 0.18{\scriptsize$\pm$.06} & 8.0 & (hindsight oracle) \\
\bottomrule
\end{tabular}
\end{table}

Table~\ref{tab:decompose} is our main result; three patterns emerge.

\emph{Lookahead is the largest single factor.} Going from reactive token-aware scaling (20.9\%) to EWMA with delay-aware lookahead (1.44\%) is a $14\times$ reduction. Without lookahead, even correct demand magnitude leaves the system structurally late by $\Delta$. This is the factor practitioners should prioritize.

\emph{Token granularity is the prerequisite.} QPS-based HPA (53\%) degrades to majority-failure, whereas token-aware reactive scaling (20.9\%) already captures the right demand magnitude. Token awareness is necessary but not sufficient; it must be paired with lookahead.

\emph{The UCB margin adds robustness at no cost; complex predictors do not pay off.} EWMA-UCB (0.54\%, 9.5 replicas) vs.\ EWMA (1.44\%, 9.9 replicas) shows the bounded margin catches tail under-predictions while the cap ($\gamma{=}0.8$) prevents over-provisioning---a $3\times$ violation reduction at no replica cost (the 0.4-replica difference is within one standard deviation). Sophistication does not help: KF-4state is \emph{worse} (5.9\%, high variance $\pm$6.9) because its constant-velocity model is mismatched to heavy-tailed bursts, and KF-6state reaches a marginally lower 0.47\% only at 17\% higher replica cost. Reactive-Plant (19\%) confirms that observing plant state without predicting ahead is insufficient.

\subsection{Cold-Start Sensitivity}

\begin{figure}[t]
\centering
\includegraphics[width=\columnwidth]{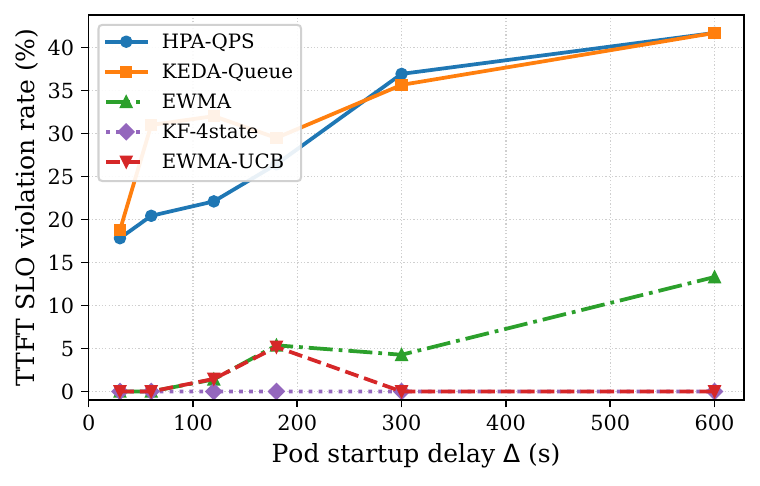}
\caption{TTFT SLO violation rate vs.\ pod startup delay $\Delta$ on the synthetic burst workload (5 seeds). Predictive-UCB methods stay near zero up to 300\,s; reactive baselines degrade monotonically with $\Delta$.}
\label{fig:coldstart}
\end{figure}

Figure~\ref{fig:coldstart} varies $\Delta$ on a 5$\to$18~req/s ramp. Reactive baselines degrade monotonically: their violation rate is a direct function of how long the system stays under-provisioned waiting for capacity. Predictive-UCB methods stay near zero across 30--600\,s, because lookahead shifts the decision point earlier by exactly $\Delta$ and the margin absorbs the larger prediction error at longer horizons. A practical consequence: one need not estimate $\Delta$ precisely---any conservative upper bound suffices. We quantify the cost of that conservatism next.

\subsection{Cost--SLO Pareto Frontier}

\begin{figure}[t]
\centering
\includegraphics[width=\columnwidth]{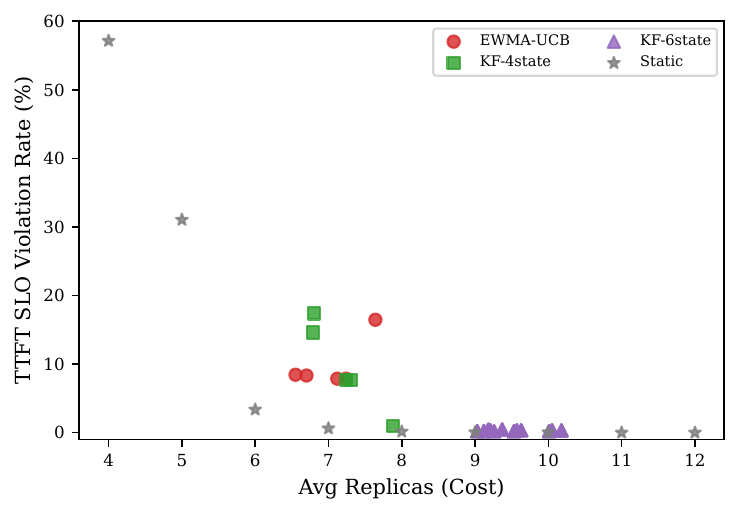}
\caption{Cost--SLO Pareto frontier on ServeGen (55 configurations across policies and parameter sweeps). Static($N$) is a hindsight-tuned lower bound, not deployable online. No dynamic method dominates Static(8); the adaptation premium is $\sim$1.5 replicas.}
\label{fig:pareto}
\end{figure}

To show the findings are not artifacts of a single parameter choice, we sweep $\beta\in\{0.5,1.0,1.5,2.0,3.0\}$ for EWMA-UCB and KF-4state and queue/KV thresholds for KF-6state (55 configurations). Figure~\ref{fig:pareto} shows the frontier. EWMA-UCB at low $\beta$ (aggressive) sits at 7--10 replicas with 3--8\% violations and moves toward the Table~\ref{tab:decompose} point as $\beta$ grows; KF-6state reaches sub-1\% only at 9+ replicas. No dynamic method dominates the hindsight-optimal Static(8). The \emph{adaptation premium}---the extra cost of scaling without prior capacity knowledge---is $\sim$1.5 replicas for EWMA-UCB (9.5 vs.\ 8.0), the fundamental price of adapting to unknown variation rather than manually right-sizing. For stable, predictable workloads, a right-sized static allocation remains more cost-efficient; the premium is justified precisely when load is uncertain.

\subsection{System Identification: What Drives Queue Buildup?}
\label{sec:sysid}

To understand \emph{why} plant-state observation helps only marginally, we apply system identification, fitting a model of the one-step queue change from current state over 1907 transitions (3 ServeGen seeds $\times$ 4 replica counts: 4, 6, 8, 10). A data-driven Ridge model attains one-step queue mean absolute error (MAE) of 0.04, versus 7.40 for the hand-written linear coupling assumed inside KF-6state---so the coupling KF-6state relies on is a poor model of the real dynamics, which is consistent with its failure to improve the tradeoff.

Single-feature ablations show queue dynamics are strongly \emph{autoregressive} ($R^2{=}0.64$ from queue history alone), while \metric{kv\_cache\_usage} and request rate individually carry almost no marginal signal ($R^2{=}0.001$ and ${-}0.004$). This does \emph{not} mean cache state is unimportant; rather, in observational traces demand, queue history, and KV usage are tightly correlated ($r{>}0.89$), so linear attribution among them is unreliable. To break the correlation we turn to controlled experiments.

\begin{table}[t]
\caption{Controlled isolation (4 replicas, single seed). At matched QPS, context length drives KV-cache pressure and TTFT far more than request rate does at short context.}
\label{tab:controlled}
\centering
\small
\begin{tabular}{@{}rrrrr@{}}
\toprule
Input Len & QPS & KV\% & Queue/rep & TTFT P95 \\
\midrule
128 & 8 & 1.9 & 0.01 & 40\,ms \\
512 & 8 & 5.2 & 0.01 & 89\,ms \\
2048 & 8 & 17.8 & 0.01 & 89\,ms \\
4096 & 8 & 34.0 & 0.13 & 693\,ms \\
\midrule
256 & 8 & 2.9 & 0.01 & 55\,ms \\
256 & 12 & 4.8 & 0.01 & 69\,ms \\
256 & 16 & 7.4 & 0.03 & 83\,ms \\
256 & 20 & 10.0 & 0.03 & 95\,ms \\
\bottomrule
\end{tabular}
\end{table}

Table~\ref{tab:controlled} provides the causal evidence the regression cannot. Holding QPS fixed at 8 and raising input length $32\times$ (128$\to$4096) raises KV usage from 2\% to 34\% and TTFT $17\times$---at identical request rate. Conversely, raising QPS $2.5\times$ at short context degrades TTFT only modestly (55$\to$95\,ms). \textbf{In systems with cache-gated admission, token weight matters more than request count}---a finding general to LLM serving. The architectural basis is vLLM's admission rule (Section~\ref{sec:method}): longer contexts consume more cache per request and trigger admission blocking sooner, regardless of GPU idle time. This is why token-aware metrics, not QPS, are the right autoscaling signal, and it explains the first jump in Table~\ref{tab:decompose}.

%% ============================================================
\section{Hardware Validation}
%% ============================================================
\label{sec:hw}

The simulator establishes \emph{relative} policy behavior; a real cluster is needed to confirm the central timing mechanism holds in the \emph{absolute} regime. We therefore validate on hardware. We are explicit about scope: this experiment validates delay-aware lookahead as the decisive control-plane mechanism, not the full token-decomposed EWMA-UCB policy or its multi-seed statistics.

\textbf{Deployment.} Qwen2.5-7B-Instruct on a Kubernetes cluster of NVIDIA A100 40\,GB GPUs (p4d.24xlarge nodes), one GPU per vLLM pod, scaling the Deployment from 1 to 4 single-GPU replicas. Key configuration:
\begin{itemize}\itemsep1pt \parskip0pt \topsep2pt
\item \texttt{tensor-parallel-size=1} (one GPU per replica);
\item \texttt{max-num-seqs=32} (matching the simulator's batch cap);
\item \texttt{gpu-memory-utilization=0.9}.
\end{itemize}
We inject a controlled 60\,s startup delay via an init-container sleep, which isolates actuation delay from unrelated image-pull and model-load variability.

\textbf{Controllers (same Deployment, 1--4 replicas).}
\emph{KEDA-Queue (reactive):} a KEDA ScaledObject scales up when \texttt{sum(vllm:\allowbreak num\_requests\_\allowbreak waiting)}${>}5$; 15\,s polling.
\emph{Predictive (lookahead):} a controller reads vLLM metrics every 15\,s, applies EWMA ($\alpha{=}0.3$) to active demand (running + waiting), estimates the slope, predicts demand at $t{+}\Delta$ ($\Delta{=}60$\,s, matching the injected delay), and adds the UCB margin ($\beta{=}1.5$, $\gamma{=}0.8$) before computing $R^\star$.

\textbf{Workload (identical for both).} 5$\to$15$\to$5 QPS over 5 minutes, mixed Pareto context lengths (256--4096), \texttt{max\_tokens=256}, driven from a co-located client.

\begin{table}[t]
\caption{Hardware results (Qwen2.5-7B, A100, vLLM v0.11, $\Delta{=}60$\,s).}
\label{tab:hw}
\centering
\small
\begin{tabular}{@{}lcccc@{}}
\toprule
Controller & Viol\% & P95 & \makecell{Scale-up\\time} & \makecell{Ready\\time} \\
\midrule
\makecell[l]{KEDA-Queue\\(reactive)} & 63.5 & 24.0\,s & $t{\approx}90$\,s & $t{\approx}150$\,s \\[3pt]
\makecell[l]{Predictive\\(lookahead)} & \textbf{3.7} & \textbf{1.6\,s} & $t{\approx}45$\,s & $t{\approx}105$\,s \\
\bottomrule
\end{tabular}
\end{table}

\textbf{Results.} Table~\ref{tab:hw} confirms the mechanism. The predictive controller detected the rising demand slope at $t{\approx}45$\,s---when active requests climbed from 0 to 19, before any queue formed---and scaled from 1 to 2 replicas; the new replica became ready at $t{\approx}105$\,s, just as the burst entered its high-QPS phase. KEDA detected queue buildup only at $t{\approx}90$\,s, and its capacity arrived at $t{\approx}150$\,s, 45\,s into the overload. The 45-second-earlier scale-up cut TTFT violations from 63.5\% to 3.7\% (P95 24.0\,s$\to$1.6\,s). The direction and magnitude match the simulator's qualitative prediction, supporting the relative-comparison claims of Section~\ref{sec:eval} while remaining honest that absolute simulator latency is off by $\sim$2$\times$.

%% ============================================================
\section{Related Work}
%% ============================================================

\textbf{LLM serving optimization.} vLLM~\cite{kwon2023vllm} introduced PagedAttention; Orca~\cite{yu2022orca} pioneered continuous batching; Sarathi-Serve~\cite{agrawal2024sarathi} chunks prefill; DistServe~\cite{zhong2024distserve} and Splitwise~\cite{patel2024splitwise} disaggregate prefill and decode; ServerlessLLM~\cite{fu2024serverlessllm} reduces cold-start latency via optimized loading. These optimize intra-node serving; we study inter-node autoscaling policy and treat the serving engine as given.

\textbf{LLM autoscaling.} llm-d~\cite{llmd2025} provides Kubernetes-native distributed inference with KV-cache-aware routing and a workload-variant autoscaler that scales prefill and decode pools independently. Chiron~\cite{patke2025chiron} combines local batch-size adaptation with global queue-based instance scaling using queue-wait estimation as an SLO proxy. Both are important advances but remain fundamentally reactive---they scale after detecting overload. Our contribution is orthogonal: a controlled analysis of \emph{which} predictive factors matter, with delay-aware lookahead identified as dominant. Recent predictive serving systems target demand forecasting directly; a head-to-head quantitative comparison under matched startup delay and workload is an important next step we leave to future work, as it requires reproducing those systems' control planes on the same cluster.

\textbf{Predictive autoscaling (general).} SHEPHERD~\cite{zhang2023shepherd} forecasts DNN-serving demand; Autopilot~\cite{rzadca2020autopilot} predicts resource needs at Google scale; classical feedback-control theory~\cite{hellerstein2004feedback} underpins such controllers. None addresses LLM-specific token costs, KV-cache dynamics, two-phase latency, or multi-minute startup delays. Our empirical finding---that for LLM autoscaling, predictor sophistication matters far less than the delay-aware UCB framework---is, to our knowledge, new.

%% ============================================================
\section{Discussion and Conclusion}
%% ============================================================

\textbf{What is Kubernetes-specific vs.\ general.} Two of our findings are general to LLM serving regardless of orchestrator: token weight dominates request count (Table~\ref{tab:controlled}), and queue buildup is gated by KV-cache pressure (Section~\ref{sec:sysid}). One is specific to the Kubernetes actuation path: the multi-minute startup delay is what makes lookahead the dominant factor (Table~\ref{tab:decompose}, Fig.~\ref{fig:coldstart}). A serving stack with fast actuation would shift the balance away from lookahead and toward demand estimation.

\textbf{Practical recommendations.} (1) Deploy EWMA-UCB first: token-aware tracking + delay lookahead + a bounded margin ($\beta{=}1.5$, $\gamma{=}0.8$) cut violations from 53\% to 0.5\%; lookahead alone is $14\times$, the margin a further $3\times$. It needs only token-rate metrics and a margin scaled by $\Delta$. (2) Monitor token throughput and \metric{gpu\_cache\_usage\_perc} together, not QPS. (3) Budget for a $\sim$1.5-replica ($\sim$19\%) adaptation premium over hindsight-optimal static; static is cheaper only when load is predictable.

\textbf{Limitations.} The simulator simplifies scheduling: absolute TTFT is $\sim$2$\times$ off Vidur, so we restrict simulator claims to relative comparisons and validate the absolute timing mechanism on hardware (Section~\ref{sec:hw}). The hardware experiment is intentionally small-scale and validates the timing mechanism, not the full token-decomposed policy or its statistics. All experiments use a single 7B model and colocated prefill/decode; larger models and disaggregated serving may exhibit different dynamics. Results use ServeGen-generated workloads; proprietary production traces would add validation.

\textbf{Conclusion.} We decomposed predictive LLM autoscaling into four factors and found that delay-aware lookahead is dominant ($14\times$), the UCB margin adds robustness at no cost, and more complex predictors yield diminishing or negative returns. Controlled experiments show token granularity is necessary because context-length-driven cache pressure---not request count---dominates TTFT degradation. A real-Kubernetes validation confirmed the mechanism: predictive lookahead cut TTFT violations from 63.5\% to 3.7\% versus reactive KEDA. The residual gap to oracle static provisioning is a scale-down control problem, not a prediction problem, motivating model-predictive control over identified plant models as future work. We will release the simulator, controllers, workloads, and plotting scripts upon publication.

\bibliographystyle{IEEEtran}
\bibliography{references}

\end{document}